# Simultaneous Imaging Achiral and Chiral Domains beyond Diffraction Limit by Structured-illumination Microscopy

*Jiwei Zhang,\* Shiang-Yu Huang, Ankit Kumar Singh, and Jer-Shing Huang\**

Dr. J. Zhang
MOE Key Laboratory of Material Physics and Chemistry under Extraordinary Conditions, and Shaanxi Key Laboratory of Optical Information Technology
School of Physical Science and Technology
Northwestern Polytechnical University
Xi'an 710129, China
E-mail: jwzhang@nwpu.edu.cn
Dr. J. Zhang, S.-Y. Huang, Dr. A. K. Singh, Dr. J.-S. Huang
Leibniz Institute of Photonic Technology
Albert-Einstein Straße 9, Jena 07745, Germany
E-mail: jer-shing.huang@leibniz-ipht.de
Dr. J.-S. Huang
Abbe Center of Photonics
Friedrich-Schiller University Jena
Max-Wien-Platz 1, 07743 Jena, Germany
Dr. J.-S. Huang
Research Center for Applied Sciences
Academia Sinica
128 Sec. 2, Academia Road, Nankang District, 11529 Taipei, Taiwan
Dr. J.-S. Huang
Department of Electrophysics
National Chiao Tung University
1001 University Road, 30010 Hsinchu, Taiwan

Modern optical microscopy methods have been advanced to provide super resolution at high imaging speed, but not chirality discriminative. We recently proposed "chiral structured-illumination microscopy (SIM)" method to image chiral fluorescent domains at sub-wavelength resolution. Chiral SIM is suitable for imaging chiral domains at sub-wavelength resolution but loses the high spatial frequency of the achiral ones. In order to obtain the full picture of all fluorescent domains at sub-wavelength resolution, we propose an advanced version of chiral SIM, termed "double SIM", which enables simultaneous imaging of achiral and chiral domains at sub-wavelength resolution. In double SIM, the illumination field must be spatially structured both in the intensity and optical chirality so that moiré effects can be concurrently generated on the achiral and chiral fluorescent domains of a sample. This allows down-modulating the high spatial frequency of both domains at the same time and thus provides sub-wavelength details after image reconstruction. To generate the illumination field with concurrently structured intensity and optical chirality, we propose interfering two coherent circularly polarized light with the same handedness at the sample plane. We introduce the working principle of double SIM and theoretically demonstrate the feasibility of this method using different kinds of synthetic samples.

**1. Introduction**

Optical microscopy has found wide applications in the research fields of biomedical imaging, device fabrication, material science, etc. The Abbe diffraction limit,[1] however, has restricted the resolution of conventional optical microscopy to around half of the illumination wavelength for more than one century. During the past decades, multiple techniques have been developed to go beyond the diffraction limit and achieve super-resolution with high imaging speed. These techniques are mainly fluorescence-detected, such as stimulated emission depletion microscopy,[2,3] stochastic optical reconstruction microscopy,[4] photoactivated localization microscopy,[5] and structured-illumination microscopy (SIM).[6,7] However, these super-resolution optical microscopy methods only detect the fluorescence emitted from the sample distribution, which are not dependent on the sample chirality.

Chirality is a pervasive property found in many biological molecules such as DNA, peptides, and proteins. Optical fields can also be chiral as the electromagnetic field rotates during the light propagation, such as left- and right-handed circularly polarized light (L-/R-CPL). To characterize the chirality of an optical field, a conservative quantity called "optical chirality (OC)" was introduced.[8,9] The differential interactions between chiral molecules and chiral optical fields lead to chiroptical effects such as circular birefringence and circular dichroism (CD). OC has been linked to the CD of chiral molecules.[9] As a well-established technique for chiral analysis, CD spectrometer has made great progress in the characterization of chiral molecules.[10] However, this method does not provide spatial resolution and thus is not suitable for microscopic investigation of the spatial distribution of chiral species. Current methods for chiral domain imaging include wide-field[11,12] and confocal CD microscopy,[13,14] second harmonic generation CD mapping,[15,16] two-photon luminescence chirality mapping,[17] photothermal CD microscopy[18,19], and chiral near-field scanning optical microscopy.[20,21] Although these methods all provide information on the spatial distribution of chiral domains, they suffer from either the diffraction-limited resolution or long image acquisition time due to the scanning nature.

Recently, we have proposed a super-resolution chiral imaging method, "chiral SIM", to fast image chiral domains at sub-wavelength resolution.[22] In typical SIM, the intensity of the illumination is spatially structured. Differently, in chiral SIM, it is the OC of the illumination being spatially structured in order to generate the moiré pattern on the distribution of chiral fluorescent domains. The corresponding high spatial frequency of the sample chirality is down-modulated and later extracted through Fourier analysis. The image of chiral domain distribution is finally obtained at sub-wavelength resolution after regular SIM image reconstruction. However, it loses the high spatial frequency of the achiral fluorescent domains because the illumination intensity is uniformly distributed and the moiré effect is only brought onto the chiral domains but not the achiral ones. Therefore, the previously proposed chiral SIM method cannot resolve the achiral fluorescent domains of the samples at sub-wavelength resolution.

In this work, we present a new super-resolution imaging method called double SIM which allows for simultaneously obtaining the super-resolution images of achiral and chiral fluorescent domains. In double SIM, the intensity and OC of the illumination fields are both spatially structured. The moiré effects are thus generated on both of the achiral and chiral domains concurrently. SIM image reconstruction is performed twice to simultaneously obtain the sub-diffraction limited images of both domains. In the following sections, we first outline the working principle of double SIM and introduce an illumination scheme by using far-field optics. Theoretical demonstrations of this method are provided by using different types of synthetic samples. The effects of shot noise on the performance of double SIM are also analyzed. Finally, we discuss the limitations of the proposed approach based on far-field optics and the possibilities of using near-field schemes to generate the illumination fields required for double SIM.

## 2. Working Principle

### 2.1. Theory

When a chiral molecule is subjected to an electromagnetic field, the absorption rate can be expressed as[9]

$$A = \frac{2}{\varepsilon_0} \left( \omega U_e \alpha'' - C G'' \right), \tag{1}$$

where $\varepsilon_0$ is vacuum permittivity and $\omega$ is the angular frequency of the field. $\alpha''$ is the imaginary electric dipole polarizability and $G''$ is the imaginary chiral polarizability. $U_e = \frac{\varepsilon_0}{4} |\mathbf{E}|^2$ and $C = -\frac{\varepsilon_0 \omega}{2} \left( \mathbf{E}^* \cdot \mathbf{B} \right)$ are the electric energy density and the OC, respectively, with $\mathbf{E}$ and $\mathbf{B}$ being the electric and magnetic field components. On the right-hand side of Equation (1), the first term relates to the dominant electric dipole absorption and the second term indicates the chirality-induced absorption. For achiral objects ($\alpha'' \neq 0, G'' = 0$), only the first term contributes to the absorption. For chiral objects ($\alpha'' \neq 0, G'' \neq 0$), the overall absorption depends on both terms.

In typical wide-field fluorescence microscopy, uniform linearly polarized illumination is usually used as the excitation light. This means $U_e(\mathbf{r})$ is a spatially invariant constant and $C(\mathbf{r}) = 0$, where $\mathbf{r}$ denotes the spatial coordinate. In this case, the image reflects the spatial distribution of all fluorescent domains of the sample $\alpha''(\mathbf{r})$ but is not chirality-dependent. The fluorescence image is given by

$$M(\mathbf{r}) = \frac{2\beta}{\varepsilon_0} \left[ \omega U_e(\mathbf{r}) \alpha''(\mathbf{r}) \right] \otimes h(\mathbf{r}), \tag{2}$$

where the symbol "$\otimes$" denotes the convolution operation, $\beta$ is a coefficient describing the imaging efficiency of the optical setup and the quantum yield of the fluorophore, and $h(\mathbf{r})$ is the point spread function (PSF) of the optical setup. To image the spatial distribution of chiral fluorescent domains, the emission intensity must be chirality-dependent. Fluorescence-detected circular dichroism (FDCD)[23–25] is a suitable method for this purpose, provided that all of the criteria of FDCD are satisfied.[24] In the wide-field FDCD method, chiral samples are sequentially excited by spatially uniform L- and R-CPL beams possessing OC ($C_{L,R}(\mathbf{r}) = \pm \frac{\varepsilon_0 \omega}{2c} |\mathbf{E}(\mathbf{r})|^2$) with opposite signs. The corresponding electric energy densities ($U_{eL,R}(\mathbf{r}) = \frac{\varepsilon_0}{4} |\mathbf{E}(\mathbf{r})|^2$) of the illumination are spatially uniform and identical for both handedness. In this case, the spatial information of chiral domains $G''(\mathbf{r})$ is calculated from the differential fluorescence image

$$\Delta M(\mathbf{r}) = M_L(\mathbf{r}) - M_R(\mathbf{r}) = \frac{2\beta}{\varepsilon_0} \left\{ \left[ C_L(\mathbf{r}) - C_R(\mathbf{r}) \right] G''(\mathbf{r}) \right\} \otimes h(\mathbf{r}), \tag{3}$$

where $M_L(\mathbf{r})$ and $M_R(\mathbf{r})$ are the fluorescence images recorded under the illumination of L- and R-CPL, respectively.

While typical wide-field fluorescence microscopy cannot distinguish chiral domains, wide-field FDCD loses the information of achiral domains after getting the differential fluorescence images. Moreover, the spatial resolution of both methods is diffraction-limited. To improve the spatial resolution, one effective method is to use SIM, which can double the maximum spatial resolution. In typical SIM, the fluorescent sample (Figure 1(a)) is illuminated by structured electric energy density pattern (Figure 1(b)). The structured illumination induces the moiré effect (Figure 1(c)), which down modulates the high spatial frequency components of fine features of the sample into the detectable frequency range of the diffraction-limited imaging system. Combining SIM with FDCD, chiral SIM structures the OC of the illumination to bring moiré effect onto the chiral domains of the sample.[22] In order to eliminate the chirality-irrelevant responses through Fourier analysis, chiral SIM employs the spatially uniform intensity of the illumination. In this way, the chirality-dependent fluorescence image has an enhanced spatial frequency bandwidth because of the modulated OC of the illumination. Consequently, the chiral domain image at sub-wavelength resolution can be reconstructed by the SIM algorithm. Details of this method can be found in our previous publication.[22] Note that this chiral SIM method requires the intensity of the illumination to be spatially uniform so that the chirality-independent part of the fluorescence can be removed in the image reconstruction. However, this operation naturally discards the achiral domain information of the samples.

To address this issue, the double SIM method proposed in this work spatially modulates both the electric energy density $U_e(\mathbf{r})$ and the OC $C(\mathbf{r})$ of the illumination fields into a sinusoidal form. To decouple the spatial information of the achiral domain ($\alpha''(\mathbf{r})$, Figure 1(a)) and the chiral domain ($G''(\mathbf{r})$, Figure 1(d)), one possible way is to produce a pair of structured illumination fields with opposite-handed OC, i.e., $C_\pm(\mathbf{r}) = \pm C_{\text{stru.}}(\mathbf{r})$ and identical structured $U_{e,\text{stru.}}(\mathbf{r})$. The opposite-handed OC patterns are displayed in Figure 1(e) and 1(f), respectively. The corresponding $U_{e,\text{stru.}}(\mathbf{r})$ patterns should not change upon flipping the handedness of the OC (Figure 1(b)). With this condition fulfilled, the moiré effects will be generated concurrently on the achiral domains (Figure 1(c)) and the chiral ones (Figure 1(g) and 1(h)). As a result, the high spatial frequency of both domains will be simultaneously down-modulated. Briefly, a pair of fluorescence images under the illumination of double SIM can be obtained as

$$M_{\pm}(\mathbf{r}) = \frac{2\beta}{\varepsilon_0}\left[\omega U_{e,\text{stru.}}(\mathbf{r})\alpha''(\mathbf{r}) \mp C_{\text{stru.}}(\mathbf{r})G''(\mathbf{r})\right] \otimes h(\mathbf{r}). \tag{4}$$

By summing up these two images,

$$M_+(\mathbf{r}) + M_-(\mathbf{r}) = \frac{4\beta}{\varepsilon_0}\left[\omega U_{e,\text{stru.}}(\mathbf{r})\alpha''(\mathbf{r})\right] \otimes h(\mathbf{r}), \tag{5}$$

the contribution from the chiral polarizability $G''(\mathbf{r})$ is removed. Therefore, the super-resolution image of achiral domains $\alpha''(\mathbf{r})$ can be reconstructed by the SIM algorithm. On the other hand, by getting the difference between these two images,

$$M_+(\mathbf{r}) - M_-(\mathbf{r}) = -\frac{4\beta}{\varepsilon_0}\left[C_{\text{stru.}}(\mathbf{r})G''(\mathbf{r})\right] \otimes h(\mathbf{r}), \tag{6}$$

the contribution from the electric dipole polarizability $\alpha''(\mathbf{r})$ is eliminated. As a result, the super-resolution image of chiral domains $G''(\mathbf{r})$ can be obtained by the SIM image reconstruction.

## 2.2. Illumination Scheme

To generate the aforementioned illumination fields for double SIM, we propose one simple yet effective illumination scheme based on far-field optics. This scheme only requires slight modification on the experimental setup of the typical SIM. As depicted in Figure 2(a), two CPL beams with identical handedness are focused on the back focal plane (BFP) of an objective. The incident angle on the sample plane is $\alpha$. The CPL beams can be described by two orthogonally polarized components with equal amplitude ($E_0$) and a phase difference of $\frac{\pi}{2}$ and $-\frac{\pi}{2}$, corresponding to L-CPL and R-CPL, respectively. In our previous work, we have systematically investigated the generation of OC patterns formed by the superposition of two plane waves in free space.[26] The resulting electric energy density and OC of the interference fields formed by two L-CPL beams and two R-CPL beams are

$$U_e(x) = \varepsilon_0 E_0^2 \left(1 - \cos^2\alpha \cos\Phi\right), \tag{7a}$$

and

$$C_\pm(x) = \pm 2\varepsilon_0 E_0^2 k_0 \left(1 - \cos^2\alpha \cos\Phi\right), \tag{7b}$$

respectively. Here $k_0$ is the wavenumber of vacuum light and $\Phi = 2k_0 x \sin\alpha + \varphi_1 - \varphi_2$ with $\varphi_{1,2}$ being the initial phase of the two CPL beams. Equation (7a) and (7b) indicate that the interference of two CPL beams with identical handedness simultaneously generates structured $U_e(x)$ and structured OC patterns. Altering the handedness of two beams changes the handedness of the structured OC pattern but not the structured intensity pattern of the illumination, which is the key requirement for the illumination fields of double SIM. We have performed rigorous numerical simulations using finite-difference time-domain method (FDTD Solutions, Lumerical) to verify the analytical solutions of Equation (7a) and (7b). The simulation results (Figure 2(b-g)) are in good agreement with the analytical solutions. As shown in Figure 2 (b), the interference of two L-CPL beams results in structured OC pattern with positive sign $C_+(x)$. Flipping the handedness of the two CPL beams to right-handed results in structured OC pattern with negative sign $C_-(x)$ (Figure 2 (c)). For both handedness, the corresponding structured $U_e(x)$ patterns remain the same regardless of the handedness of the two CPL beams (Figure 2(d) and 2(e)). The line-cut profiles are displayed in Figure 2(f) and 2(g), respectively. The detailed simulation method can be found in our previous work.[26]

## 3. Results and Discussion

### 3.1. Theoretical demonstration

*3.1.1. Siemens star sample*

In this section, we theoretically demonstrate the enhanced resolving power and the discriminability of double SIM for both achiral and chiral fluorescent domains. The sample is a synthetic Siemens star divided into four quadrants with different combinations of the following compositions, namely non-fluorescent background ($\alpha''(\mathbf{r})=0, G''(\mathbf{r})=0$), achiral fluorescent domain ($\alpha''(\mathbf{r})\neq 0, G''(\mathbf{r})=0$), and left- and right-handed fluorescent chiral domains ($\alpha''(\mathbf{r})\neq 0, G''(\mathbf{r})\neq 0$), as illustrated in Figure 3 (a). The corresponding spatial distribution of achiral domains ($\alpha''(\mathbf{r})$) and chiral domains ($G''(\mathbf{r})$) are shown in Figure 3(b) and 3(c), respectively. In the following, we will show that both of the achiral and chiral domain images at sub-wavelength can be obtained simultaneously by double SIM.

The illumination scheme in Figure 2(a) is used in the theoretical demonstration. In order to set reasonable simulation parameters of the sample and the illumination fields, we consider the dimensionless dissymmetry factor which is commonly used in chiral analysis. This factor is defined by the chirality-dependent absorption difference over the average absorption, namely $g_{\mathrm{CPL}} \equiv 2\frac{A_\mathrm{L} - A_\mathrm{R}}{A_\mathrm{L} + A_\mathrm{R}}$, where $A_\mathrm{L}$ and $A_\mathrm{R}$ is the absorption rate of the sample under the illumination of L- and R-CPL, respectively. The magnitude of $g_{\mathrm{CPL}}$ for conventional chiral molecules is small, ranging from $10^{-2}$ to $10^{-5}$. In our theoretical demonstrations, the $|g_{\mathrm{CPL}}|$ of the sample is set to be $10^{-3}$. The illumination fields are numerically simulated by the FDTD method, the simulation of the raw image acquisition and the SIM image reconstructions are performed with MATLAB (R2017a). The operational procedure is summarized in the Supporting Information.

For a fair comparison, the typical wide-field fluorescence image and the wide-field FDCD image are simulated with the spatially uniform illumination of L- and R-CPL. Figure 3(d-g) present the demonstration results of the wide-field imaging and double SIM. To show a clear resolution improvement provided by double SIM, the noise effect, which will be discussed in the next section, is not considered in this simulation. The typical wide-field fluorescence image and wide-field FDCD image at diffraction-limited resolution reveal the information of achiral and chiral domain distributions, as shown in Figure 3(d) and 3(e), respectively. In contrast, the super-resolution achiral and chiral domain images obtained by double SIM are displayed in Figure 3(f) and 3(g), respectively. The reduced unresolved central regions in both of the achiral and chiral domain images from the double SIM clearly indicate an enhanced resolution because of the expanded detectable bandwidth of the optical imaging system. It's worthy to point out that the super-resolution achiral domain image (Figure 3(f)), which can also be obtained by the typical SIM method, can neither distinguish between achiral and chiral fluorescent domains nor identify the handedness of chiral fluorescent domains (lower half of Figure 3(a)). On the other hand, the super-resolution chiral domain image (Figure 3(g)), which can also be obtained by our previous chiral SIM method, is blind to the achiral fluorescent domains (upper-left quadrant of Figure 3(a)), despite the remarkable chiral discriminability at sub-wavelength resolution. As a result, the upper-left quadrant of the sample appears to be structureless in the super-resolution chiral domain image shown in Figure 3(g). In other words, simultaneously resolving achiral and chiral domains in a complex sample at sub-wavelength resolution can only be realized by the double SIM method presented in this work but not by typical SIM or chiral SIM.

*3.1.2. Nanobead sample*

In this section, we theoretically demonstrate the ability of double SIM using nanobeads (diameter of 75, 100 or 150 nm) to mimic a real sample. The beads are made of either an achiral material or enantiomers of a chiral material with $g_{\text{CPL}} = 10^{-3}$. Compared to the typical wide-field fluorescence image (Figure 4(a)) and wide-field FDCD image (Figure 4(b)), the super-resolution images obtained by double SIM on achiral (Figure 4(c)) and chiral domains (Figure 4(d)) possess a higher spatial resolution, respectively. To quantitatively compare the spatial resolution, we additionally set several bead pairs (diameter of 75 nm) with different separation distances in the simulated sample. Two bead pairs with separation distances of 100 nm and 125 nm are located along the same horizontal line in the field of view, as marked by the dashed lines in Figure 4(a-d). Figure 4(e) and 4(f) display the corresponding line-cut profiles obtained from the wide-field microscopy images and double SIM images. The results show that the beads separated by 125 nm, which are originally unresolved in the wide-field images (see red curves in Figure 4(e) and 4(f)), are now well resolved in the double SIM images (see green curves in Figure 4(e) and 4(f)). For the beads separated by shorter distances, *e.g.*, 100 nm, even the double SIM method cannot resolve the beads clearly. This result agrees with the theoretical resolution of double SIM, which is around 115 nm with the current illumination scheme. In theory, the double SIM has a resolution enhancement with a highest factor of ~2 over the wide-field imaging method. Detailed analysis on the theoretical resolution of double SIM is in Supporting Information.

This demonstration result also shows that the handedness of the chiral domains which cannot be identified in the achiral domain image (Figure 4(c)), now can be distinguished as the red and blue colors in the chiral domain image at sub-wavelength resolution (Figure 4(d)). In addition, we set an achiral bead pair in the simulated sample, as marked by the dashed ellipses in Figure 4(a-d). The results show that this achiral bead pair which are not resolved in the chiral domain image (Figure 4(d)), are now well resolved in the achiral domain image (Figure 4(c)). To summarize, only with the proposed double SIM method, the super-resolution images of the achiral and chiral domains can be obtained simultaneously, which is favorable when investigating samples with complex domain distributions.

**3.2 Noise Effects**

In practice, noise that occurs during the raw image acquisition may affect the double SIM image reconstruction. As for the illumination scheme in Figure 2(a) based on far-field optics, the noise is usually dominated by the shot noise. In this section, we present a theoretical demonstration of double SIM that takes into account the effect of shot noise. The simulated sample is a Siemens star formed by left-handed domain embedded in the background with right-handedness. The noise package of the MATLAB simulation used for the raw image acquisition is from the DIPimage toolbox.[27] Compared to the typical wide-field fluorescence image (Figure 5(a)), the fluorescence image at sub-wavelength resolution obtained by double SIM (Figure 5(b)) exhibits a better spatial resolution as expected. However, both of the wide-field FDCD image (Figure 5(c)) and the chiral domain image obtained by double SIM (Figure 5(d)) are quite noisy and the resolution enhancement is not very pronounced. The reason lies in that the CD response of the chiral sample is intrinsically weak. The fluorescence modulation induced by the structed OC is thus small and buried in the noise during the raw image acquisition. This finally degrades the quality of the reconstructed chiral domain image.

For SIM image reconstruction, it requires the fluorescence modulation depth to be large enough to overcome the background noise. Thus, we evaluate the ratio of the fluorescence modulation depth to the noise $R$ in the real space. Because the chirality-induced absorption is much smaller than the electric dipole absorption, $R$ for the chiral domain image reconstruction ($R_{chiral}$) is thus nearly three orders of magnitude smaller than that for the achiral domain image reconstruction ($R_{achiral}$). As a result, the noise effect is more pronounced in the chiral domain image reconstruction than that in the achiral domain image reconstruction. Detailed discussion can be found in the Supporting Information.

To suppress the noise effect, enhancement of the modulation depth-to-noise ratio $R$ is necessary. Possible strategies include increasing the illumination power, extending the raw image acquisition time or developing new illumination schemes with well-designed plasmonic[28–30] or dielectric nanostructures.[31–34] Because the shot noise is dominate during the raw image acquisition, one of the practical strategies to enhance the quality of the reconstructed chiral domain image is using image averaging with numbers of raw images.[35] Figure 5(e) and 5(f) show the reconstructed wide-field FDCD image and the chiral domain image at sub-wavelength resolution after averaging one thousand raw images. One can observe that both of the chiral domain images are less noisy compared to those in Figure 5(c) and 5(d), respectively.

Importantly, the resolution improvement of the chiral domain image obtained by double SIM over the wide-field FDCD image is clearer. In addition, as the number of raw images for image averaging increases, the noise effect becomes less pronounced (see Supporting Information).

In practical experiments, optical components such as the dichroic mirror and objective can distort the polarization state of the input CPL into elliptically polarized light (EPL). This polarization distortion reduces the modulation of the structured illumination and thus reduces the modulation depth-to-noise ratio during the raw image acquisition. Therefore, imperfection of the circular polarization leads to the degraded image quality (see Supporting Information). In experiment, strategies to maintain high degree of circular polarization, *e.g.*, using orthogonally positioned identical dichroic mirrors and strain-free objective, should be carefully introduced. This is particularly important for samples with small chiral dissymmetry.

## 4. Conclusion and Outlook

We have proposed and demonstrated the double SIM that allows for simultaneously obtaining achiral and chiral domain images at sub-wavelength resolution. In double SIM, the illumination scheme that provides both structured OC and intensity patterns is required. In addition, the intensity patterns should remain unchanged while altering the handedness of the OC patterns. Super-resolution images of fluorescent achiral and chiral domain distributions of complex samples can be simultaneously obtained since the moiré effects are generated on both domains. Based on the far-field optics, we have proposed one possible illumination scheme using the interference of two coherent circularly polarized beams with the same handedness and described the corresponding operational procedure. We have demonstrated theoretically double SIM using different kinds of samples with quantitative consideration on the noise effect and resolving power. We also show that the weak CD signal directly leads to low quality of the reconstructed chiral domain image and using raw image averaging technique can effectively enhance the modulation-to-noise ratio. As the resolution improvement of double SIM is governed by the spatial frequency of the structured illumination, the maximum resolution improvement over the uniform illumination-based method is only two when using the illumination scheme based on far-field optics.

To further enhance the spatial resolution, one of the potential strategies is to exploit the evanescent waves (EWs) or surface plasmon waves (SPWs) because their wavelengths are shorter than that of the far-field excitation light. The periodic patterns formed by the interference of EWs[36–38] and SPWs[39–42] are much finer. This has been utilized to enhance the resolution in typical SIM. On the other hand, well-designed plasmonic and dielectric nanostructures may also provide fine illumination patterns with high spatial frequency determined by the nanostructure geometry. The resolution of typical SIM has been further improved using this type of illumination approach via the blind SIM algorithm.[43–50] This solution, however, is more challenging for double SIM as double SIM requires that the intensity pattern stays the same while changing the handedness of the local optical filed nearby the nanostructures. The difficulty lies in the fact that as nanostructures are illuminated with CPL, the spin-orbit coupling is almost inevitable.[51,52] As a result, the near-field intensity pattern always varies when the handedness of the illuminating CPL is changed. Smart designs of the nanostructures are required to suppress the variation of the field intensity distribution due to the spin-orbit coupling. The proposed double SIM may find applications in the characterization of biological chiral targets, such as collagen and DNA, and the analysis of drug, polymer, or chiral inorganic nanostructures.

**Supporting Information**

**Acknowledgements**
The support from the DFG (HU2626/3-1, HU2626/6-1 and CRC 1375 NOA) and the NSFC (62005219) is acknowledged. J. Zhang acknowledges the support from Sino-German (CSC-DAAD) Postdoc Scholarship Program, 2018. We thank R. Heintzmann for providing the SIM reconstruction code.

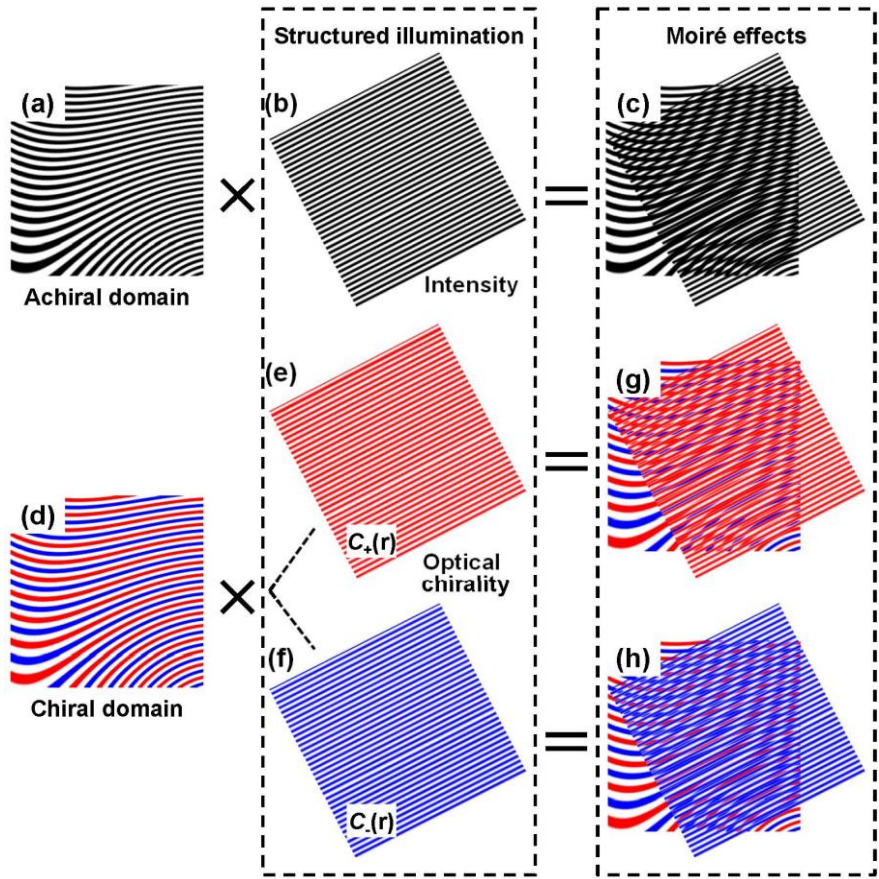

**Figure 1.** (a) Achiral domain of the sample. (b) Structured electric energy density pattern. (c) Moiré pattern generated on the achiral domain in double SIM. The white and black color denote the achiral domain of the sample and the intensity of the illumination. (d) Chiral domain of the sample. Structured OC patterns with (e) left-handedness and (f) right-handedness. (g, h) Moiré patterns generated on the chiral domain in double SIM. The red and blue color denote the chiral domain of the sample and the OC of the illumination with opposite handedness. The moiré patterns possessing lower spatial frequency than that of the sample.

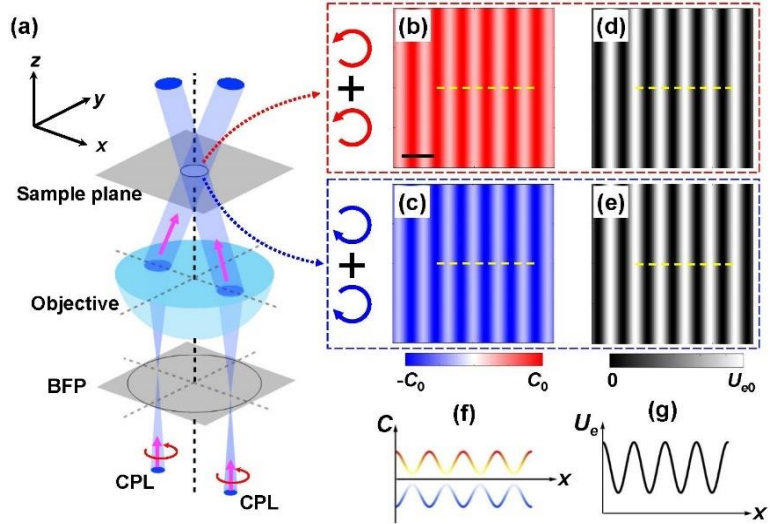

**Figure 2.** Illustration of the illumination scheme of double SIM using far-field optics. (a) Schematic of one possible approach to generate the illumination patterns with structured electric energy density and OC simultaneously. Structured OC patterns formed by the interference of (b) two L-CPL beams $C_+(x)$ and (c) two R-CPL beams $C_-(x)$. (d, e) Structured patterns of $U_e(x)$. The OC patterns in (b) and (c) possess the opposite handedness. The $U_e$ patterns in (d) and (e) are the same. Line-cut profile of the (f) OC and (g) $U_e$ along the dashed lines in (b-e). Scale bar in (b) is $2\pi/k_0$ and applicable for (d-e).

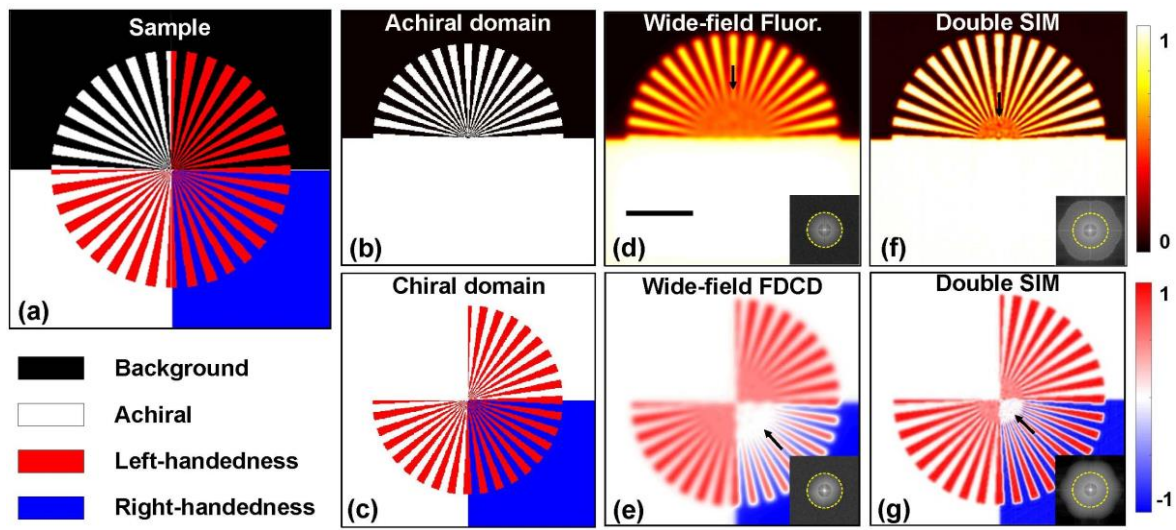

**Figure 3.** Theoretical demonstration of double SIM using a synthetic Siemens star formed by four different materials. (a) Designed sample with four combinations (from left to right and top to bottom): "non-fluorescent background (black color) and achiral fluorescent domain (white color)", "non-fluorescent background and chiral fluorescent domain", "achiral and chiral fluorescent domains" and "chiral fluorescent domains with opposite-handedness (red and blue color denotes the left- and right-handedness, respectively)". These combinations represent all the possible domain distributions of complex samples. (b) Achiral $\alpha''(\mathbf{r})$ and (c) chiral domain distribution $G''(\mathbf{r})$ corresponding to the sample in (a). Simulated (d) typical wide-field fluorescence image and (e) wide-field FDCD image at diffraction-limited resolution. Simulated (f) achiral and (g) chiral domain images at sub-wavelength resolution obtained by double SIM. Scale bar in (d) is 2 μm and applicable for (e-g). Color bar of (d) and (f) indicates normalized fluorescence. Color bar of (e) and (g) indicates normalized differential fluorescence. The arrows in (d-g) mark the unresolved regions resulted from the diffraction limit. Insets in (d-g) are the Fourier transformations of each image with the dashed circles indicating the bandwidth of the optical transfer function of the diffraction-limited imaging system.

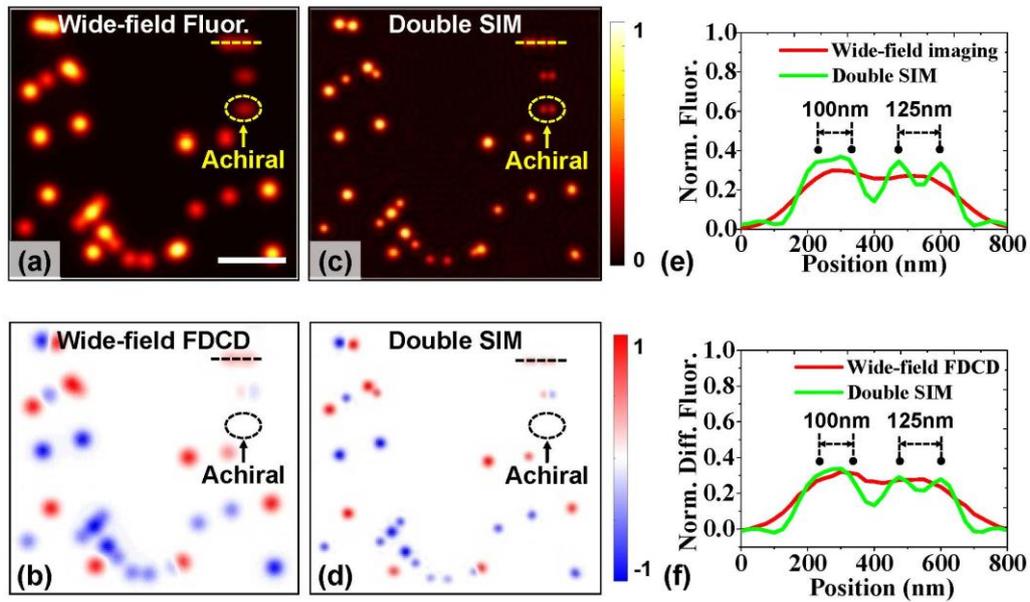

**Figure 4.** Theoretical demonstration of double SIM using randomly distributed nanobeads. Simulated (a) typical wide-field fluorescence image and (b) wide-field FDCD image at diffraction-limited resolution. Simulated (c) achiral and (d) chiral domain images at sub-wavelength resolution obtained by double SIM. Scale bar in (a) is 1 μm and applicable for (b-d). Color bar of (a) and (c) indicates normalized fluorescence. Color bar of (b) and (d) indicates normalized differential fluorescence. (e) Line-cut profiles of the wide-field fluorescence image (red curve) and achiral domain image at sub-wavelength resolution (green curve) along the yellow dashed lines in (a) and (c). (f) Line-cut profiles of the wide-field FDCD image (red curve) and chiral domain images at sub-wavelength resolution (green curve) along the black dashed lines in (b) and (d). The black dots in (e) and (f) represent the locations of the two bead pairs along the horizontal dashed lines in (a-d). The bead separation distances are indicated in (e) and (f).

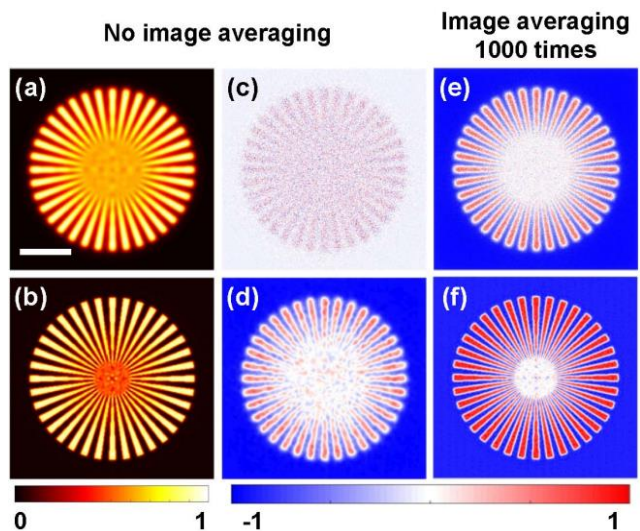

**Figure 5.** Theoretical demonstration of double SIM with the consideration of noise effect. Simulated (a) typical wide-field fluorescence image at diffraction-limited resolution, (b) sub-wavelength achiral domain image, (c) wide-field FDCD image at diffraction-limited resolution, and (d) sub-wavelength chiral domain image when taking into account the noise during the raw image acquisition. Reconstructed (e) wide-field FDCD image and (f) sub-wavelength chiral domain image with averaging one thousand raw images. Color bar of (a) and (b) indicates normalized fluorescence. Color bar of (c-f) indicates normalized differential fluorescence. Scale bar in (a) is 2 μm and applicable for (b-f).

# Supporting Information

**Simultaneous Imaging Achiral and Chiral Domains beyond Diffraction Limit by Structured-illumination Microscopy**

*Jiwei Zhang, Shiang-Yu Huang, Ankit Kumar Singh, and Jer-Shing Huang\**

**S.1 Operational procedure of double SIM based on far-field optics**

The operational procedure of double SIM is depicted in Fig. S.1. In the first step (Fig. S.1(a)), two L-CPL beams and two R-CPL beams are successively focused on the back focal plane (BFP) of the objective in order to generate the structured illumination patterns discussed in the main text. Upon this illumination, the sample with both achiral and chiral domains emits fluorescence and the raw images are acquired. Similar to the typical SIM, the structured patterns of the illumination in double SIM are created in three in-plane orientations with three phases for the isotropic resolution enhancement. The phase can be adjusted by controlling the phase difference between the two incident beams and the in-plan orientation can be adjusted by rotating the grating or grating image on a spatial light modulator. For each orientation and phase, a couple of fluorescence raw images $M_+(\mathbf{r})$ and $M_-(\mathbf{r})$ are recorded under the illumination patterns with the opposite-handed OC $C_\pm(\mathbf{r})$ and the identical electric energy density $U_e(\mathbf{r})$. Overall, eighteen raw images are acquired. In the second step (Fig. S.1(b)), through the subtraction and addition of each couple of raw images, two groups of nine sub-images are generated to be processed by the SIM algorithm. Finally, the super-resolution images of achiral and chiral domains can be reconstructed.

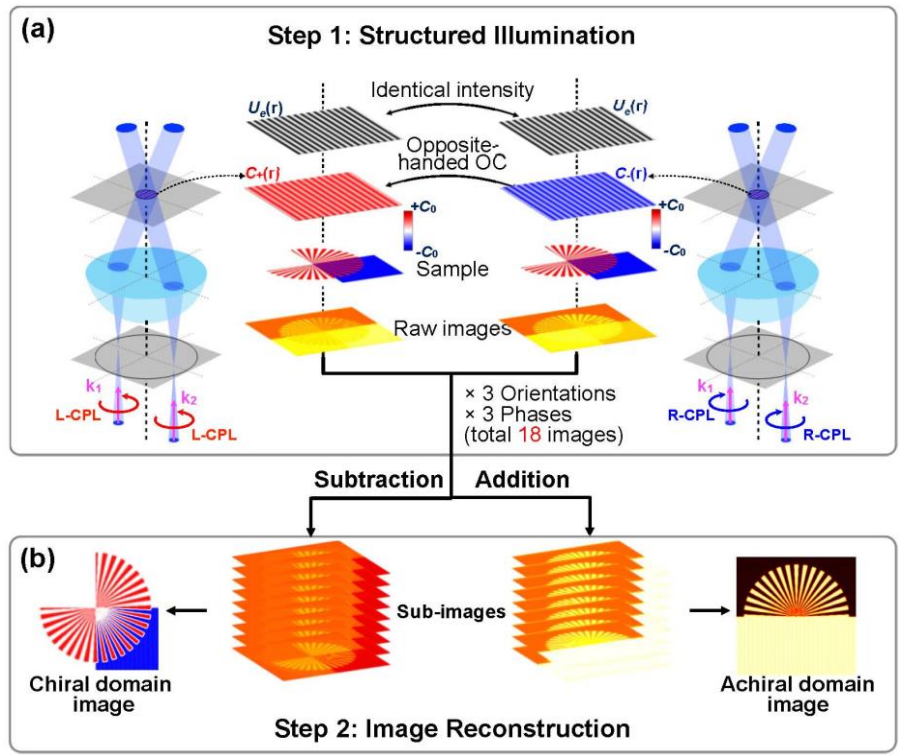

**Figure S.1.** Illustration of the operational procedure in double SIM using far-field optics.

## S.2 Theoretical resolution of double SIM

Considering using the same objective for the illumination and fluorescence detection, the spatial resolution of double SIM can be expressed as

$$\delta = \frac{2\pi}{k_{cutoff} + |\mathbf{k}|}, \quad (S1)$$

where $k_{cutoff} = 4\pi NA/\lambda_{em}$ is the cutoff frequency of the diffraction-limited imaging setup with NA being the numerical aperture of the objective and $\lambda_{em}$ the emission wavelength. $\mathbf{k}$ represents the wavevector of the structured $U_e$ or OC. For the structured patterns generated by using the illumination scheme in Figure 2(a), Equation (7) indicates that the absolute value of the wavevectors of $U_e$ and OC patterns are $k_{U_e} = k_C = 2k_0 n \sin\alpha$, $n$ is the refractive index of the surrounding media. Since $k_0 = \frac{2\pi}{\lambda_{ex}}$ ($\lambda_{ex}$ is the excitation wavelength) and the numerical aperture of the imaging system $NA = n \sin\alpha_{max}$, the maximum of $k_{U_e}$ and $k_C$ are determined by $k_{max} = \frac{4\pi NA}{\lambda_{ex}}$. As $\lambda_{em}$ is very close to $\lambda_{ex}$, the highest spatial resolution improvement of double SIM over the wide-field imaging method is $(k_{cutoff} + k_{max})/k_{cutoff} \sim 2$, which is the same as that of typical SIM. As for the simulation parameters of our theoretical demonstration $\lambda_{ex} = 405\text{nm}$, $NA = 1.2$ and $\alpha = 41°$, the $|\mathbf{k}| = 0.0204\text{nm}^{-1}$, $k_{cutoff} = 0.0342\text{nm}^{-1}$, $|\mathbf{k}|/k_{cutoff}$ is around 0.6. The final resolution of double SIM is around 115 nm.

## S.3 Noise effects

Taking into account the noise, the recorded signal on one camera pixel can be expressed as[22]

$$S(\mathbf{r})=\mu(\mathbf{r})+N(\mathbf{r})=\frac{2\beta}{\varepsilon_0}\left[\omega U_e(\mathbf{r})\alpha''-C(\mathbf{r})G''\right]\Delta t\otimes h(\mathbf{r})+N(\mathbf{r}), \quad (S2)$$

where $\mu(\mathbf{r})$ is the expected signal, $\Delta t$ is the image acquisition time, and $N(\mathbf{r})$ is the total background noise. To investigate the fluorescence modulation from the structured illumination, we assume that the sample is homogeneous with uniform spatial distribution of $\alpha''$ and $G''$. During the raw image acquisition, the noise is usually dominated by the shot noise $N_S$, i.e., $N\approx N_S$. Since the photon receiving process of the camera is random, the shot noise of the individual raw image can be described by one standard deviation of the Poisson distribution, which is $Std[S(\mathbf{r})]=Std[N(\mathbf{r})]=\sqrt{\mu(\mathbf{r})}$. Here, $Std$ denotes the standard deviation. After applying the structured illuminations generated using the scheme in Figure 2(a) in the main text, the sub-images are generated for achiral and chiral domain image reconstruction by obtaining the sum ($M_+(\mathbf{r})+M_-(\mathbf{r})$) and difference ($M_+(\mathbf{r})-M_-(\mathbf{r})$) of the raw images taken with left- and right-handed structured OC. Combining Equation (5-7) and Equation (S2), the signal on one pixel of the sub-images for achiral ($S_{\text{achiral}}(\mathbf{r})$) and chiral domain image reconstruction ($S_{\text{chiral}}(\mathbf{r})$) are

$$S_{\text{achiral}}(\mathbf{r})=4\beta\omega\alpha'' E_0^2\left(1-\cos^2\alpha\cos\Phi\right)\Delta t\otimes h(\mathbf{r})+N_+(\mathbf{r})+N_-(\mathbf{r}), \quad (S3a)$$

and

$$S_{\text{chiral}}(\mathbf{r})=8\beta G'' E_0^2 k_0\left(1-\cos^2\alpha\cos\Phi\right)\Delta t\otimes h(\mathbf{r})+N_+(\mathbf{r})-N_-(\mathbf{r}), \quad (S3b)$$

where the term $\cos\Phi$ indicates that the fluorescence signals are modulated due to the structured illumination, $N_+(\mathbf{r})$ and $N_-(\mathbf{r})$ are the noise of the raw image taken with left- and right-handed structured OC. Due to the addition and subtraction of the raw images, the noise of the sub-images for the achiral and chiral domain image reconstruction can be described by evaluating the standard deviation of the sub-images

$$Std[S_{achiral}(\mathbf{r})] = \sqrt{Var[S_{achiral}(\mathbf{r})]}, \tag{S4a}$$

and

$$Std[S_{chiral}(\mathbf{r})] = \sqrt{Var[S_{chiral}(\mathbf{r})]}, \tag{S4b}$$

where $Var$ denotes the variance. Because (i) the first terms on the right-hand sides of Equation (S3a) and (S3b) are temporally invariable, (ii) the noise $N_+(\mathbf{r})$ and $N_-(\mathbf{r})$ are independent with each other (*i.e.*, the covalence of $N_+(\mathbf{r})$ and $N_-(\mathbf{r})$ is zero), and (iii) $Var[N_\pm(\mathbf{r})] = \mu_\pm(\mathbf{r})$ as the shot noise follows the Poisson distribution, Equation (S4a) and (S4b) become

$$\begin{aligned} Std[S_{achiral}(\mathbf{r})] &= \sqrt{Var[N_+(\mathbf{r}) + N_-(\mathbf{r})]} \\ &= \sqrt{Var[N_+(\mathbf{r})] + Var[N_-(\mathbf{r})]} \\ &= \sqrt{\mu_+(\mathbf{r}) + \mu_-(\mathbf{r})} \\ &= 2E_0\sqrt{\beta\omega\alpha''(1-\cos^2\alpha\cos\Phi)\Delta t \otimes h(\mathbf{r})}, \end{aligned} \tag{S5a}$$

and

$$\begin{aligned} Std[S_{chiral}(\mathbf{r})] &= \sqrt{Var[N_+(\mathbf{r}) - N_-(\mathbf{r})]} \\ &= \sqrt{Var[N_+(\mathbf{r})] + Var[N_-(\mathbf{r})]} \\ &= \sqrt{\mu_+(\mathbf{r}) + \mu_-(\mathbf{r})} \\ &= 2E_0\sqrt{\beta\omega\alpha''(1-\cos^2\alpha\cos\Phi)\Delta t \otimes h(\mathbf{r})}, \end{aligned} \tag{S5b}$$

when considering the scaling property of the variance. Here, $\mu_+(\mathbf{r})$ and $\mu_-(\mathbf{r})$ are the expected signals on one pixel of the raw images taken with left- and right-handed structured OC, respectively. Equation (S5a) and (S5b) indicate that the noise of the sub-images for achiral and chiral domain image reconstruction are the same, regardless of the raw image adding or subtracting.

The fluorescence modulation depth, *i.e.*, the difference between the maximun and minimun expected signals of the sub-images for achiral ($D_{achiral}$) and chiral domain image reconstruction ($D_{chiral}$) can be derived from Equation (S3a) and (S3b) as

$$D_{achiral} = 8\beta\omega\alpha'' E_0^2 \cos^2\alpha \Delta t \otimes h, \tag{S6a}$$

and

$$D_{chiral} = 16\beta G'' E_0^2 k_0 \cos^2\alpha \Delta t \otimes h. \tag{S6b}$$

The ratio of the fluorescence modulation depth to the noise is defined as $R \equiv \dfrac{D}{Std[S(\mathbf{r})]}$. Therefore, combining Equation (S5) and (S6), the modulation depth-to-noise ratio for achiral ($R_{achiral}$) and chiral domain image reconstruction ($R_{chiral}$) are given by

$$R_{achiral} = 4E_0 \sqrt{\beta\omega\alpha'' \frac{\cos^4\alpha}{1-\cos^2\alpha\cos\Phi} \Delta t \otimes h}, \tag{S7a}$$

and

$$R_{chiral} = 8E_0 \sqrt{\beta \frac{G''^2 k_0^2 \cos^4\alpha}{\omega\alpha''(1-\cos^2\alpha\cos\Phi)} \Delta t \otimes h}. \tag{S7b}$$

Equation (S7a) and (S7b) show that both ratios are determined by the imaging efficiency of the optical setup and the quantum yield of the fluorophore ($\beta$), the acquisition time ($\Delta t$), the PSF of the imaging system, and the illumination field ($\alpha$, $E_0$, $\omega$). The ratio between $R_{chiral}$ and $R_{achiral}$ is

$$\eta = \frac{R_{chiral}}{R_{achiral}} = 2\frac{G''}{c\alpha''}, \tag{S8}$$

where, $c$ is the vacuum light speed. Considering $g_{CPL} = -\dfrac{2C_{CPL}G''}{\omega U_{e,CPL}\alpha''}$ and $|g_{CPL}|$ is set to be $10^{-3}$ in the simulation, the ratio of $\dfrac{G''}{\alpha''} = 1.5\times 10^5$ (SI Units). As a result, $R_{chiral}$ is three orders of magnitude smaller than $R_{achiral}$.

**S.4 Double SIM image quality improved by raw image averaging**

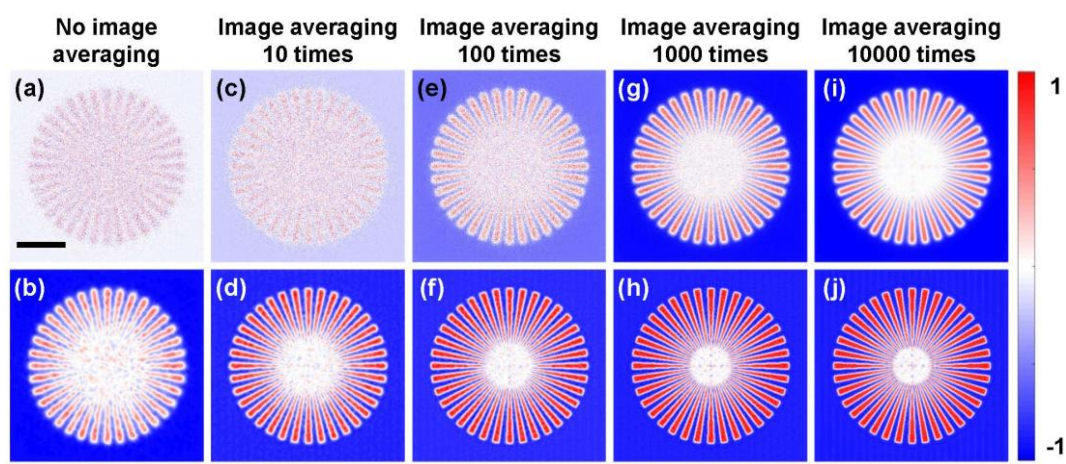

**Figure S.2.** Demonstration results of the chiral domain images when different numbers of raw images are used for image averaging. Simulated (a) chiral domain image at diffraction-limited resolution and (b) sub-wavelength resolution obtained by double SIM with no image averaging. (c, e, g, i) and (d, f, h, j) are the results after averaging ten, one hundred, one thousand, and ten thousand raw images. Color bar indicates normalized differential fluorescence. Scale bar in (a) is 2 μm and applicable for (b-j).

## S.5 Structured patterns of the interference between two EPL beams

We evaluate the structured illumination patterns generated by the interference of two EPL beams. In the first scenario, we consider that the EPL beam is comprised by two orthogonally polarized components with the equal amplitudes of $E_0$ and a phase difference of $\theta$, where $\theta$ determines the ellipticity of the EPL. As a result, the structured electric energy density $U_e$ remains unchanged, *i.e.*, the same as Equation (7a) in the main text. However, the structured OC of Equation (7b) becomes

$$C_\pm(x) = \pm 2\varepsilon_0 E_0^2 \sin\theta k_0 \left(1 - \cos^2\alpha \cos\Phi\right). \tag{S9}$$

For CPL where $\theta = \pm\frac{\pi}{2}$, Equation (S9) is the same as Equation (7b) in the main text. Equation (S9) indicates that the contrast of the structured OC pattern, *i.e.*, the difference over the sum of the maximum and minimum value, $\gamma = \cos^2\alpha$ remains the same as that obtained by the interference of two pure CPL beams. However, as $|\sin\theta| < 1$, the amplitude of the structured OC formed by the interference of two EPL beams is smaller than that formed by two CPL beams and determined by the ellipticity. The decreased OC will induce a weak CD signal and the noise effect will get more pronounced in the chiral domain image reconstruction.

In the second scenario, unequal amplitudes and $\pm\frac{\pi}{2}$ phase difference between the two orthogonally polarized components can also result in an EPL beam. By controlling the ratio between the amplitudes of the two components, EPL beams with variable ellipticity can be obtained. We simulated the corresponding structured $U_e$ and OC patterns formed by the interference of two EPL beams. The simulation results also show a reduced OC amplitude with an unchanged contrast. Meanwhile, the contrast of the structured $U_e$ decreases as the ellipticity of the EPL beams increases (Figure S.3). This will induce small fluorescence modulation depth-to-noise ratio for achiral domain image reconstruction.

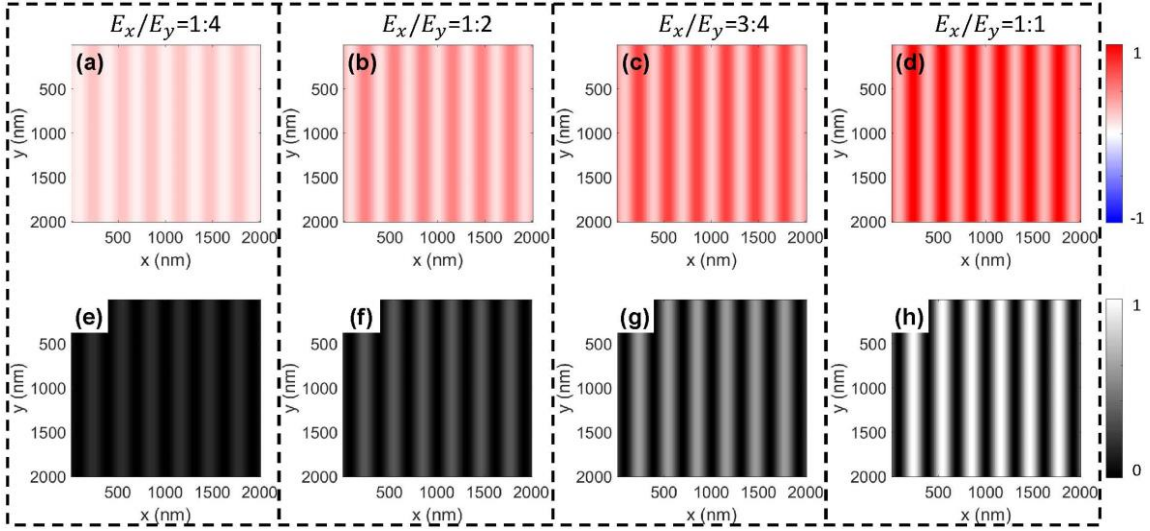

**Figure S.3.** Simulated structured (a-d) OC and (e-h) $U_e$ patterns formed by the interference of two EPL beams with variable ellipticity. Here, the EPL is comprised by two orthogonally polarized components with unequal amplitudes $E_x$, $E_y$ and a phase difference of $\pi/2$. The ratio of $E_x / E_y$ determines the ellipticity of the EPL.